\documentclass[twoside]{article}
\usepackage{fleqn,espcrc2}

\usepackage{epsfig}
\usepackage[figuresright]{rotating}

\newcommand{\AmS}{{\protect\the\textfont2
  A\kern-.1667em\lower.5ex\hbox{M}\kern-.125emS}}

\title{Spin-charge separation: From one hole to finite doping}
\author{Z.Y. Weng, D.N. Sheng, and C.S. Ting \address{Texas Center for Superconductivity,
University of Houston, Houston, TX 77204, U.S.A.}}
\begin{document}
\begin{abstract}

In the presence of nonlocal phase shift effects, a quasiparticle can remain 
topologically stable even in a spin-charge separation state due to the
confinement effect introduced by the phase shifts at finite doping.  True 
{\em deconfinement} only happens in the {\em zero-doping} limit where a bare
hole can lose its integrity and decay into holon and spinon elementary
excitations. The Fermi surface structure is completely different in these
two cases, from a large band-structure-like one to four Fermi points in
one-hole case, and we argue that the so-called underdoped regime actually
corresponds to a situation in between.
\vspace{1pc}
\end{abstract}
\maketitle
Spin-charge separation idea \cite{anderson} has become a very basic concept in 
understanding the doped Mott-Hubbard insulator models related to 
high-$T_c$ 
cuprates. It states that the system has two independent elementary excitations, 
neutral spinon and spinless holon, respectively, as opposed to a single 
elementary excitation in conventional metals --- the quasiparticle that 
carries both spin and charge quantum numbers. In literature, it has been usually 
assumed that the quasiparticle excitation would no longer be stable in such a 
spin-charge separation state and should decay into more elementary spinon 
and holon excitations once being created, e.g., by injecting a bare hole into 
the system as illustrated in Fig. 1(a).

But in this work we will show an example that the fate of the quasiparticle 
can be drastically different from what has been previously perceived. Specifically, based on a consistent spin-charge separation theory we find that a 
quasiparticle excitation does {\em not} simply break up into a pair of free 
holon and spinon and disappear in the metallic regime. On the contrary, it 
remains topologically {\it stable} in a form of spinon-holon bound state whose 
symmetry is fundamentally different from a pair of simple spinon and holon
as sketched in Fig. 1(b).

Such a spin-charge separation theory \cite{string1,string3} is based on a 
generalized slave-particle representation of the $t-J$ model in which the 
electron (hole) operator is expressed as follows 
\begin{equation}
c_{i\sigma }=h_i^{\dagger }b_{i\sigma }e^{i\hat{\Theta}_{i\sigma }}
\label{mutual}
\end{equation}
where holon $h_{i }^{\dagger }$ and spinon $b_{i\sigma }$ are both
{\em bosonic} fields, satisfying the no-double-occupancy constraint $%
h_i^{\dagger }h_i+\sum_\sigma b_{i\sigma }^{\dagger }b_{i\sigma }=1$. These 
spinon and holon fields describe the elementary excitations in the mean-field 
state \cite{string3} which is controlled by a single bosonic RVB order 
parameter 
$\Delta^s$ and reduces to the Schwinger-boson mean-field state \cite{aa} in the 
limit of half-filling. 

The key feature in the bosonization formulation (\ref{mutual}) is the phase
shift field $e^{i\hat{\Theta}_{i\sigma }}\equiv (-\sigma )^ie^{i\Theta _{i\sigma
}^{string}}$ with $\Theta _{i\sigma }^{string}\equiv \frac 1 2\left[ \Phi
_i^b-\sigma \Phi _i^h\right] $, in which $\Phi _i^b=\sum_{l\neq i}\theta
_i(l)\left( \sum_\alpha \alpha n_{l\alpha }^b-1\right) $, $\Phi
_i^h=\sum_{l\neq i}\theta _i(l)n_l^h$, and $\theta _i(l)=\mbox{Im ln
$(z_i-z_l)$}$. It is a vortex-like phase field which precisely keeps track of
the singular phases (signs) involved in the doped $t-J$ model known as the phase string effect \cite{string1}.
\begin{figure}[ht!]
\epsfxsize=4.6 cm
\centerline{\epsffile{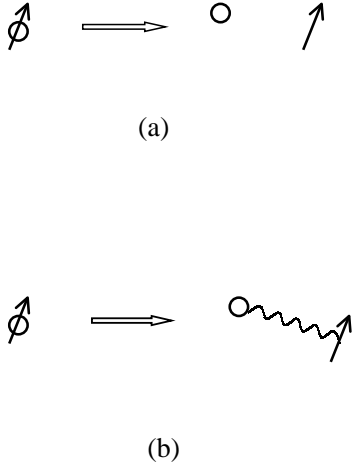}}
\vspace{0.5mm}
\caption{Quasiparticle deconfinement (a) and 
confinement (b) due to the phase shift field} 
\label{fig:1}
\end{figure}

The decomposition (\ref{mutual}) implies that in order for a quasiparticle created 
by $c_{i\sigma }$ to decay into holon-spinon elementary excitations, a phase 
shift $\hat{\Theta}_{i\sigma}$ must take place in the background. Since 
$\hat{\Theta}_{i\sigma}$ is basically an {\it infinite-body} operator in the
thermodynamic limit, the dissolution of a quasiparticle would take virtually 
infinite time to realize under a local perturbation which means 
that a quasiparticle will practically remain a stable and independent excitation 
even in a background of holon-spin sea. In fact, it can be
shown \cite{string5} that $e^{i\hat{\Theta}_{i\sigma }}$, due to its vorticities, 
introduces a new ``angular-momentum-like'' quantum number to the quasiparticle
such that the excitation is orthogonal to the ground state as well 
as those states of elementary holon and spinon excitations. Therefore, the
phase shift field not only plays a role leading to a non-Fermi liquid (with the 
quasiparticle weight $Z=0$) as conjectured \cite{anderson,anderson1} by Anderson 
on a general ground, but also attaches a distinct quantum number to the 
quasiparticle to ensure its integrity in the metallic phase.

To further understand the phase shift effect, let us construct 
$|\Psi ^{\prime }\rangle \equiv e^{i\hat{\Theta}_{i\sigma }}|\Psi _G\rangle $ where $|\Psi 
_G\rangle $ is the ground state of $H_{t-J}$. Then by computing its energy cost
one can show \cite{string5} 
\begin{equation}
\langle \Psi ^{\prime }|H_{t-J}|\Psi ^{\prime }\rangle -\langle \Psi
_G|H_{t-J}|\Psi _G\rangle \propto \ln L  \label{E'}
\end{equation}
which diverges logarithmically with the sample length scale $L$. But the 
quasiparticle state $c_{i\sigma }|\Psi_G\rangle $ should cost only a finite 
energy relative to the ground-state energy. It thus imples that the holon and 
spinon constituents on the r.h.s. of (\ref{mutual}) can no longer behave as 
mean-field free elementary excitations: They have to absorb the effect of the 
vortex-like phase shift $e^{i\hat{\Theta}_{i\sigma }}$ to result in the 
finiteness of the quasiparticle energy. Clearly here one has to go beyond
the mean-field treatment of individual holon and spinon as 
$c_{i\sigma}$ involves an infinite body of them. Upon detailed examination, it 
is found \cite{string5} that the holon field $h_i^{\dagger }$ has to be bound to 
$e^{\frac {i}{2}\Phi _i^b}$ while $b_{i\sigma }$ to $e^{-i\frac \sigma 
2\Phi_i^h}$ in (\ref{mutual}). But while $e^{i\hat{\Theta}_{i\sigma }}$ is
well-defined, these two phase factors are not single-valued by themselves
except in the zero doping limit. Consequently, at finite doping the holon
and spinon constituents are generally confined together due to their binding
to $e^{i\hat{\Theta}_{i\sigma }}$.

Involving {\em infinite-body} holons and spinons, the quasiparticle excitation has to be treated
more carefully {\it before} applying the mean-field approximation. Using the
{\it exact} Hamiltonian to determine the equation of motion $-i\partial_tc_{i
\sigma}(t)=[H_{t-J}, c_{i\sigma}(t)]$ and {\it then} introducing the mean-field
order parameter $\Delta ^s$ in the high order terms, one finds \cite{string5}
\begin{eqnarray}\label{cc}
-i\partial _tc_{i\sigma }(t) \approx t_{eff}\sum_{l=NN(i)}c_{l\sigma 
}+\mu c_{i\sigma }  \nonumber  \\
- \frac 38J\Delta^s \sum_{l=NN(i)}\left[ e^{i\hat{%
\Theta}_{i\sigma }}h_i^{\dagger }b_{l-\sigma }^{\dagger }e^{i\sigma
A_{il}^h}+ ... \right]
\end{eqnarray}
where $t_{eff}=\frac t2(1+\delta )$, $\mu$ is the chemical potential. $A_{il}^h$
is the difference of $\Phi^h_i/2$ at site $i$ and $l$. Note that without the ``scattering'' term, (\ref{cc}) would give rise to a bare quasiparticle kinetic 
energy $\epsilon _{{\bf k}}=-2t_{eff}(\cos k_x+\cos k_y)$. The
``scattering'' effect in the second line of (\ref{cc}) actually corresponds to 
the decaying process of the quasiparticle into holon and spinon constituents.
But the presence of $e^{i\hat{\Theta}_{i\sigma}}$ prevents a real decay since 
such a vortex field would cost a logarithmically divergent energy if being left 
alone as discussed above. Nonetheless, at 
\begin{equation}\label{energy}
E_{\mbox{quasiparticle}}>E_{\mbox{holon}}+E_{\mbox{spinon}}  
\end{equation}
a strong signature of spin-charge separation is expected to show up in the 
spectral function as the logarithmic potential is not important in short-range,
high-energy regime. 

In the ground state, the bosonic holons are Bose condensed with $%
\langle h_i^{\dagger}\rangle= h_0\sim \sqrt{\delta}$ and the d-wave 
superconducting
order parameter $\Delta^{SC}\neq 0$ \cite{string3}. Based on (\ref{cc}), one can find a coherent contribution from the ``scattering'' term. The decomposition 
(\ref{mutual}) may be rewritten in two parts: $
c_{i\sigma}= h_0 {a}_{i\sigma}+ {c}_{i\sigma}^{\prime}$
where ${a}_{i\sigma}\equiv b_{i\sigma}e^{i\hat{\Theta}_{i\sigma}}$ and ${c}%
^{\prime}_{i\sigma}=(:h_i^{\dagger}:)b_{i\sigma} e^{i\hat{\Theta}_{i\sigma}}$
with $: h_i^{\dagger}:\equiv h_i^{\dagger}- h_0$. Correspondingly, the 
single-electron propagator may be expressed as 
\begin{equation}
G_e= h_0^2{G}_a+{G}_e^{\prime },  \label{g12}
\end{equation}
where one has \cite{string5}
\begin{equation}
h_0^2{G}_a({\bf k},\omega )\sim h_0^2\left( \frac{u_{{\bf k}}^2}{\omega -E_{%
{\bf k}}}+\frac{v_{{\bf k}}^2}{\omega +E_{{\bf k}}}\right) .  \label{barg}
\end{equation}
with $E_{{\bf k}}=\sqrt{(\epsilon_{{\bf k}}-\mu)^2+|\Delta_{{\bf k}}|^2}$,
$\Delta_{{\bf k}}=\frac 3 4 J\sum_{{\bf q}} \Gamma_{{\bf q}}\left( \frac{%
\Delta^{SC}_{{\bf k+q}}}{h_0^2}\right)$ in which $\Gamma_{{\bf q}}=\cos q_x 
+\cos q_y $. And like in the BCS theory, $u_{%
{\bf k}}^2=[1+ (\epsilon_{{\bf k}}-\mu)/E_{{\bf k}}]/2$ and $v_{{\bf k}%
}^2=[1- (\epsilon_{{\bf k}}-\mu)/E_{{\bf k}}]/2$.

The large ``Fermi surface'' is defined by $\epsilon_{{\bf k}}=\mu$ and $%
\Delta_{{\bf k}}$ then represents the energy gap opened at the Fermi
surface which is d-wave-like as illustrated in Fig. 2 by the ``V'' shape lines.
Note that in the ground state, there also exists an independent discrete spinon 
excitation level \cite{string3} at $E_s\sim \delta J$ which leads to $E_g=2E_s\sim 41 $$meV$ (if $J\sim 100$$meV$) magnetic peak at $\delta \sim 
0.14$. This latter spin collective mode 
is {\it independent} of the quasiparticle excitation at the mean-field level. 
\begin{figure}[b!]
\epsfxsize=4.7 cm
\centerline{\epsffile{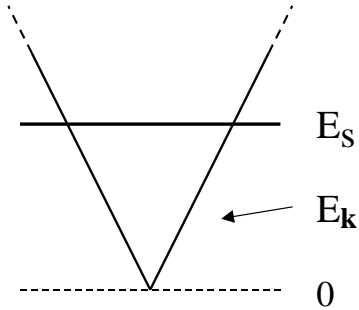}}
\vspace{0.5mm}
\caption{Low-lying excitations in the superconducting phase: The ``V"" shape
quasiparticle spectrum and the discrete spinon energy at $E_s$. }
\label{fig:4}
\end{figure}
While $\Delta^{SC}_{{\bf k}}$ apparently scales with the doping
concentration $\delta$ ($h_0 \propto \sqrt {\delta}$) and vanishes at $%
\delta \rightarrow 0$, the gap $\Delta_{{\bf k}}$ does
not and can be {\em extrapolated} to a finite value in the zero doping
limit where $T_c=0$. Namely
\begin{equation}  \label{pred1}
\frac{2\Delta_{{\bf k}}(T=0)}{T_c}\rightarrow \infty
\end{equation}
at $\delta\rightarrow 0$, whereas the BCS theory predicts a constant. 

The quasiparticle thus gains a ``coherent'' part $h_0a$ which should
behave similarly to the conventional quasiparticle in the BCS theory as it
does not further decay at $E_{{\bf k}}<E_s$. In this sense, the
quasiparticle partially restores its coherence in the superconducting state.
Such a coherent part will disappear as a result of vanishing superfluid
density. The total spectral function as the imaginary part of $G_e$
can be written as $
A_e({\bf k},\omega )=h_0^2{A}_a({\bf k},\omega )+{A}_e^{\prime }({\bf k}%
,\omega )$ so that at $h_0\rightarrow 0$, even though $\Delta _{{\bf k}}$ does 
not scale with $h_0$, the superconducting coherent part $h_0^2A_a$ vanishes
altogether, with $A_e$ reduces to the normal part $A_e^{\prime }$ at $T>T_c$.

{\it Destruction of Fermi surface: Deconfinement of quasiparticle.}
The existence of a large Fermi surface, coinciding with the {\em %
non-interacting} band-structure one, can be attributed to the integrity of
the quasiparticle due to the confinement of spinon and holon. But such a 
confinement disappear in the zero-doping limit where $\Phi^h_i$ vanishes while
$e^{\frac {i}{2}\Phi _i^b}$ becomes single-valued. In this limit, the 
single-electron propagator may 
be expressed in the following {\em decomposition} form 
\begin{equation}  \label{decomp}
G_e \approx i G_f\cdot G_b
\end{equation}
where 
\begin{equation}  \label{gh}
G_f=-i\left\langle T_t h^{\dagger}_i(t)\left( e^{i\frac 1 2 {\Phi}%
^b_{i}(t)}e^{-i\frac 1 2 {\Phi}^b_{j}(0)}\right) h_j(0)\right\rangle
\end{equation}
and 
\begin{equation}  \label{gb}
G_b =-i(-\sigma)^{i-j}\left\langle T_t b_{i\sigma}(t)b^{%
\dagger}_{j\sigma}(0)\right\rangle
\end{equation}
without the multi-value problem. 

Here the large Fermi surface is gone except for four {\em Fermi points} at ${\bf 
k}_0=(\pm\pi/2, \pm\pi/2)$ with the rest part looking like all 
``gapped'' \cite{1hole}. In fact, in the convolution form
of (\ref{decomp}) the ``quasiparticle'' peak (edge) is essentially
determined by the spinon spectrum $E^s_{{\bf k}}=2.32J\sqrt{1-s_{%
{\bf k}}^2}$ with $s_{{\bf k}}=(\sin k_x+ \sin k_y)/2$ through the
spinon propagator $G_b$, since the holon propagator $G_f$ is incoherent \cite
{1hole}. The intrinsic broad feature of the spectral function is obtained due 
to the convolution law of (\ref{decomp}) and is a
direct indication of the composite nature of the quasiparticle, which is
also consistent with the ARPES results \cite{arpes1} as well as the earlier
theoretical discussion \cite{laughlin1}. Furthermore, at high energy or 
equal-time limit, the singular phase factor 
$e^{\frac {i}{2}\Phi _i^b}$ also contributes to a large
``remnant Fermi surface'' in the momentum distribution function $n({\bf k})$
which can be regarded as a precursor of the large Fermi surface in the
confining phase at finite doping, which is also consistent with the ARPES
experiment \cite{arpes1}.

The above one-hole picture may have an important implication for the
so-called pseudo-gap phenomenon in the underdoped region of the
high-$T_c$ cuprates. Even though the confinement will set in once the
density of holes becomes finite, the ``confining force'' should remain {\em %
weak} at small doping, and one expects the virtual ``decaying'' process in (%
\ref{cc}) to contribute significantly at weakly doping to bridge a continuum
evolution between the Fermi-point structure in the zero-doping limit to a
full large Fermi surface at larger doping. Recall that in the one-hole case
decaying into spinon-holon composite happens around ${\bf k}_0$ at zero
energy transfer, while it costs {\em higher} energy near $(\pi, 0)$ and $%
(0,\pi)$, which shouldn't be changed much at weakly doping. In the
confinement regime, the quasiparticle peak in the electron spectral function
defines a quasiparticle spectrum and a large Fermi surface as discussed
before. Then due to the the virtual ``decaying'' process in the equation of
motion (\ref{cc}), the spectral function will become
much broadened with its weight shifted toward higher energy like a
gap-opening near those portions of the Fermi surface far away from ${\bf k}%
_0 $, particularly around four corners $(\pm\pi, 0)$ and $(0,\pm\pi)$. With
the increase of doping concentration and reduction of the decaying effect,
the suppressed quasiparticle peak can be gradually recovered starting from
the inner parts of the Fermi surface towards four corners $(\pm\pi, 0)$ and $%
(0, \pm\pi)$. 

Furthermore, at small doping (underdoping), something more dramatic can
happen in the present spin-charge separation state \cite{string3}: a microscopic type of {\em phase separation} has been
found in this regime which is characterized by the Bose condensation of
bosonic spinon field. Since spinons are presumably condensed in {\em %
hole-dilute} regions \cite{string3}, the propagator will then exhibit
features looking like in an even {\em weaker} doping concentration or more
``gap'' like than in a uniform case, below a characteristic temperature $%
T^{*}$ which determines this microscopic phase separation. Therefore, the
``spin gap'' phenomenon related to the ARPES experiments in the
underdoped cuprates may be understood as a ``partial'' deconfinement of
holon and spinon whose effect is ``amplified'' through a microscopic phase
separation in this weakly-doped regime. $T^{*}$ also characterizes other 
``spin-gap''properties in magnetic and transport channels in this underdoping 
regime \cite{string3}.

\end{document}